\documentclass[useAMS,usenatbib]{mn2e}
\usepackage{txfonts} 
\usepackage{epstopdf}
\usepackage[dvips]{graphicx}

\usepackage{hyperref}
\usepackage[usenames,dvipsnames]{color}
\definecolor{menublue}{rgb}{0.0,0.0,0.5}
\definecolor{citegreen}{rgb}{0.0,1.0,0.0}
\definecolor{urlred}{rgb}{1.0,0.0,0.0}
\hypersetup{bookmarksopen,pdfstartview={FitH},colorlinks=true, breaklinks=true,menucolor=menublue,urlcolor=urlred,citecolor=citegreen,linkcolor=blue}
\usepackage[all]{hypcap}

\def\del#1{{}}

\newcommand{\ltsima}{$\; \buildrel < \over \sim \;$}
\newcommand{\lsim}{\lower.5ex\hbox{\ltsima}}
\newcommand{\gtsima}{$\; \buildrel > \over \sim \;$}
\newcommand{\gsim}{\lower.5ex\hbox{\gtsima}}
\newcommand{\bra}{\langle}
\newcommand{\ket}{\rangle}

\newcommand{\dd}{\mathrm{d}}

\newcommand{\ci}{\mathrm{i}}

\newcommand{\vecl}{\bmath{L}}
\newcommand{\trace}{\mathrm{tr}}

\newcommand{\dirac}{\delta_D}

\newcommand{\vect}{\bmath{\theta}}

\newcommand{\lhat}{\hat{\bmath{L}}}
\newcommand{\vecq}{\bmath{q}}
\newcommand{\barq}{\bar{\vecq}}

\title[likelihood of intrinsic alignments]
{Parameter likelihood of intrinsic ellipticity correlations}
\author[F. Capranico, Ph.M. Merkel, B.M. Sch{\"a}fer]{
Federica Capranico$^1$\thanks{capranico@ari.uni-heidelberg.de}, Philipp M. Merkel$^2$, Bj{\"o}rn Malte Sch\"afer$^1$\\
$^1$Astronomisches Recheninstitut, Zentrum f{\"u}r Astronomie, Universit{\"a}t Heidelberg, M{\"o}nchhofstra{\ss}e 12, 69120 Heidelberg, Germany\\
$^2$Institut f{\"ur} theoretische Astrophysik, Zentrum f{\"u}r Astronomie, Universit{\"a}t Heidelberg, Albert-Ueberle-Stra{\ss}e 2, 69120 Heidelberg, Germany}

\begin{document}
\pagerange{\pageref{firstpage}--\pageref{lastpage}}
\pubyear{2012}
\maketitle
\label{firstpage}

\begin{abstract}
Subject of this paper are the statistical properties of ellipticity alignments between galaxies evoked by their coupled angular momenta. Starting from physical angular momentum models, we bridge the gap towards ellipticity correlations, ellipticity spectra and derived quantities such as aperture moments, comparing the intrinsic signals with those generated by gravitational lensing, with the projected galaxy sample of EUCLID in mind. We investigate the dependence of intrinsic ellipticity correlations on cosmological parameters and show that intrinsic ellipticity correlations give rise to non-Gaussian likelihoods as a result of nonlinear functional dependencies. Comparing intrinsic ellipticity spectra to weak lensing spectra we quantify the magnitude of their contaminating effect on the estimation of cosmological parameters and find that biases on dark energy parameters are very small in an angular-momentum based model in contrast to the linear alignment model commonly used. Finally, we quantify whether intrinsic ellipticities can be measured in the presence of the much stronger weak lensing induced ellipticity correlations, if prior knowledge on a cosmological model is assumed.
\end{abstract}


\section{Introduction}
Weak cosmic shear, i.e. lensing by the gravitational field of the cosmic matter distribution \citep{1991MNRAS.251..600B, 1994CQGra..11.2345S, 1994A&A...287..349S, 1998MNRAS.301.1064K}, is considered to be an excellent probe of structure formation processes, precision measurements of cosmological parameters \citep{1999ApJ...522L..21H, 2002PhRvD..66h3515H, 2002PhRvD..65b3003H, 2004ApJ...601L...1T, 2006JCAP...06..025H} and the influence of dark energy on cosmic structure formation \citep{2001PhRvD..64l3527H, 2002PhRvD..65f3001H, 2010GReGr..42.2177H, 2011MNRAS.413.1505A, 2012arXiv1204.5482K}. The primary observable are ellipticity correlation functions or their Fourier counterparts \citep{1997ApJ...484..560J, 1999ApJ...514L..65H, 2001ApJ...554...67H, 2004PhRvD..70d3009H}. These shape correlations have been first detected by a number of research groups more than 10 years ago \citep{2000A&A...358...30V, 2000astro.ph..3338K, 2000MNRAS.318..625B, 2000Natur.405..143W} and are now routinely used for parameter estimation. Correlations in shapes of galaxies are introduced because light rays from neighbouring galaxies experience correlation distortions due to correlations in the tidal fields through which the respective rays propagate. A common assumption is the absence of intrinsic correlations such that any positive shape correlation can be attributed to the gravitational lensing effect. This hypothesis, however, might be flawed as there are physical mechanisms by which galaxies are intrinsically shape correlated: Due to the fact that neighbouring galaxies form from correlated initial conditions, their respective angular momenta are correlated \citep{2000ApJ...545..561C, 2000MNRAS.319..649H, 2001ApJ...559..552C, 2002MNRAS.332..788M}. Assuming that the galactic disks are established with their symmetry axes colinear with the host haloes' angular momentum directions one would observe galaxies at correlated angles of inclination and therefore with correlated ellipticities.

A possible consequence of this new source of ellipticity correlation is its interference with the determination of cosmological parameters from weak lensing data, in particular the properties of dark energy. This issue has been the target of a number of investigations: Commonly, the description of intrinsic ellipticity correlations was based on the linear alignment model \citep{2001MNRAS.320L...7C, 2004PhRvD..70f3526H},
\begin{equation}
\epsilon_+ = C \left(\partial_x^2-\partial_y^2 \right)\Phi 
\qquad \mathrm{and} \qquad 
\epsilon_{\times} = 2C\: \partial_x\partial_y\Phi.
\label{eqn_linear_alignment}
\end{equation}
which provides a direct modelling of the ellipticity field on the tidal shears $\partial_\alpha\partial_\beta\Phi$ (here, the $z$-axis of the coordinate system is aligned with the line-of-sight) and is able to give a consistent description of gradient and vorticity modes of the ellipticity field. The constant of proportionality was fixed by comparison with observations \citep{2007NJPh....9..444B, 2012arXiv1203.6833J}.

If galaxy ellipticities are in fact described by an alignment model linear in the tidal fields, cosmological parameters, in particular the dark energy equation of state parameters would be severly biased \citep{2007NJPh....9..444B, 2010A&A...523A...1J, 2010MNRAS.408.1502K, 2011arXiv1109.4536K}. Apart from ellipticity correlations themselves, ellipticity position-correlations were affected and ellipticity data would exhibit cross-correlations between intrinsic ellipticities and weak lensing \citep[see, in particular,][]{2004PhRvD..70f3526H}.

There are basically four ways of dealing with intrinsic aligments. Firstly, they can be removed from data by using the fact that they are a small scale phenomenon \citep{2003MNRAS.339..711H, 2005A&A...441...47K,2002A&A...396..411K, 2003A&A...398...23K} which takes place at the cost of increasing statistical uncertainties. Secondly, one can take advantage of the fact that intrinsic alignments have different statistical properties in comparison to weak lensing ellipticity correlations \citep{2002ApJ...568...20C, 2003A&A...398...23K}, most notably it is possible to use the statistics of vortical excitations in the ellipticity field which are exclusively sourced by intrinsic alignments. Thirdly, one can design line-of-sight weightings that null out contributions due to intrinsic alignments \citep{2005A&A...441...47K, 2009A&A...507..105J,2008A&A...488..829J} which marginally increase statistical uncertainties on cosmological parameters. Parameter inference from spectra that result from data in this way still yields unbiased estimates. Finally, one can parameterise the intrinsic alignment contribution to weak lensing data and have those model parameters be determined by data alongside the cosmological parameters under consideration. Maginalisation over the parameters entering the intrinsic alignment model then propagates the statistical errors of the alignment model on to the cosmological model. With a physically correct alignment model the estimates of cosmological parameters will remain unbiased. The feasibility of this approach under the assumption of Gaussian likelihoods has been demonstrated \citep{2007NJPh....9..444B, 2009ApJ...695..652B, 2010A&A...523A...1J, 2010MNRAS.408.1502K, 2012MNRAS.423.1750L}.

The motivation of this work was to explore intrinsic alignment effects and their observable properties in angular momentum-based alignment models. In these models, the ellipticity is quadratic in the tidal shear field and because they use in principle a mechanical model of angular momentum generation and ellipticity alignment, the model parameters can be constrained from information other than ellipticity data. We will need two physically meaningful variables: a parameter which is related to the angular momentum model and whose value can be measured in structure formation simulations and a disk morphology parameter which is accessible in galaxy surveys. Clearly, quadratic alignment models will differ in their prediction of ellipticity correlations compared to linear alignment models. Together with the above results employing linear alignment models we hope to complete the view on intrinsic alignments and their relevance for future weak lensing surveys.

The aim of this paper is threefold: (i) We investigate and compare two angular-momentum based alignment models in their predictions for ellipticity correlations and formulate these predictions in terms of ellipticity correlation functions, ellipticity spectra and the scale-dependence of the variance of the ellipticity fields and compare these predictions with the equivalent quantities sourced by weak gravitational lensing (Sects.~\ref{sect_theory} and~\ref{sect_ellipticity}). (ii) The dependence of the two ellipticity models in consideration on cosmological parameters is investigated and their likelihoods are derived. With this knowledge, we quantify the contamination of weak lensing data with an intrinsic alignment contribution and quantify how this contaminations impacts on the estimation of cosmological parameters (Sect.~\ref{sect_likelihood}). (iii) We investigate if there is a possibility of observing intrinsic correlations in the presence of much stronger lensing-induced ellipticity correlations and develop statistical methods for answering these questions (Sect.~\ref{sect_lensing}). Throughout we will focus on intrinsic ellipticity correlations caused by correlated angular momenta, which is an applicable model for spiral galaxies. Those intrinsic alignments are proportional to the squared tidal field, in contrast to the linear alignment model valid for elliptical galaxies. In this limit, we neglect cross-correlations between intrinsic ellipticity alignments with the tidal field and gravitational lensing, as those correlations are proportional to the expectation value of the tidal field cubed, which vanishes in the case of Gaussian statistics. Specifically, we consider the case of EUCLID's weak lensing survey \citep{2012arXiv1206.1225A}.

The reference cosmological model used is a spatially flat $w$CDM cosmology with Gaussian adiabatic initial perturbations in the cold dark matter density field. The parameter choice is motivated by the WMAP7 results \citep{2011ApJS..192...18K,2011ApJS..192...16L}: $\Omega_m = 0.25$, $n_s = 1$, $\sigma_8 = 0.8$, $\Omega_b=0.04$ and $H_0=100\: h\:\mathrm{km}/\mathrm{s}/\mathrm{Mpc}$, with $h=0.72$. The dark energy equation of state is set to $w=-0.95$.

\section{cosmology}\label{sect_theory}

\subsection{Dark energy cosmologies}
In spatially flat Friedmann-Lema{\^i}tre Robertson-Walker cosmologies with the matter density parameter $\Omega_m$ and a dark energy component with equation of state $w(a)$, the Hubble function $H(a)=\dd\ln a/\dd t$ is given by
\begin{equation}
\frac{H^2(a)}{H_0^2} = \frac{\Omega_m}{a^{3}} + (1-\Omega_m)\exp\left(3\int_a^1\dd\ln a\:(1+w(a))\right).
\end{equation}
The value $w\equiv -1$ corresponds to the cosmological constant $\Lambda$. The Hubble function describes the time evolution of the metric and can be used for relating comoving distance $\chi$ and scale factor $a$:
\begin{equation}
\chi = c\int_a^1\dd a\:\frac{1}{a^2 H(a)},
\end{equation}
in units of the Hubble distance $\chi_H=c/H_0$. The Hubble function also determines the critical density, $\rho_\mathrm{crit} \equiv 3H^2 / (8\pi G)$.

\subsection{CDM power spectrum}
The linear CDM density power spectrum $P(k)$ describes the fluctuation amplitude of the Gaussian homogeneous density field $\delta$,
\begin{equation}
\bra\delta(\bmath{k})\delta(\bmath{k}^\prime)\ket=(2\pi)^3\dirac(\bmath{k}+\bmath{k}^\prime)P(k), 
\end{equation}
and is given by the ansatz $P(k)\propto k^{n_s}T^2(k)$ with the transfer function $T(k)$. In cosmologies with low $\Omega_m$, $T(k)$ is fitted by \citep{1986ApJ...304...15B, 1995ApJS..100..281S}:
\begin{equation}
T(q) = \frac{\ln(1+2.34q)}{2.34q}\left(1+3.89q+(16.1q)^2+(5.46q)^3+(6.71q)^4\right)^{-\frac{1}{4}}.
\label{eqn_cdm_transfer}
\end{equation}
The wave number $k$ is rescaled with the shape parameter $\Gamma\simeq\Omega_mh$, $q = k/\Gamma$. The spectrum $P(k)$ is normalised to the variance $\sigma_8$ of the density field on scales of $R=8~\mathrm{Mpc}/h$,
\begin{equation}
\sigma^2_R 
= \frac{1}{2\pi^2}\int\dd k\:k^2 P(k) W^2(kR)
= \int\dd\ln k\: \Delta^2(k) W^2(kR).
\end{equation}
$W(x)=3j_1(x)/x$ is the Fourier-transformed spherical top hat filter function. $j_\ell(x)$ refers to the spherical Bessel function of the first kind of order $\ell$ \citep{1972hmfw.book.....A, 2005mmp..book.....A} and the dimensionless variance per logarithmic wavenumber $\Delta^2(k)=k^3P(k)/(2\pi^2)$ can be used instead of the CDM spectrum $P(k)$. In computing ellipticity correlation functions and ellipticity spectra we will employ a smoothed CDM spectrum $P(k) \rightarrow P(k)\exp(-(kR)^2)$ with a smooothing scale $R$ that corresponds to a mass cutoff at a halo mass $M$. Those two quantities are related by $M = 4\pi/3\:\Omega_m\rho_\mathrm{crit}\:R^3$.

\subsection{Linear structure growth}
As long as the amplitudes in the cosmic density field are small, $\delta\ll1$, the density field grows in a homogenous way, $\delta(\bmath{x},a)=D_+(a)\delta(\bmath{x},a=1)$. The growth function $D_+(a)$ results from solving the growth equation \citep{1997PhRvD..56.4439T, 1998ApJ...508..483W, 2003MNRAS.346..573L},
\begin{equation}
\frac{\dd^2}{\dd a^2}D_+(a) + \frac{1}{a}\left(3+\frac{\dd\ln H}{\dd\ln a}\right)\frac{\dd}{\dd a}D_+(a) = 
\frac{3}{2a^2}\Omega_m(a) D_+(a).
\label{eqn_growth}
\end{equation}
Nonlinear structure formation enhances the CDM-spectrum $P(k,a)$ on small scales by one and a half order of magnitude, which is described by the fit suggested by \citet{2003MNRAS.341.1311S}. 

\subsection{Angular momenta from tidal shearing}
Angular momenta of dark matter haloes embedded in potential flows in the large-scale structure are generated by a mechanism refered to as tidal shearing, where the differential motion of mass elements inside a protohalo gives rise to a torquing moment \citep{1949MNRAS.109..365H, sciama, 1969ApJ...155..393P, 1970Afz.....6..581D, 1984ApJ...286...38W}:
\begin{equation}
L_\alpha = a^3 H(a)\frac{\dd D_+}{\dd a}\epsilon_{\alpha\beta\gamma}I_{\beta\delta}\Phi_{\delta\gamma},
\label{eqn_tidal_shearing}
\end{equation}
i.e. it is the variation $\partial_\alpha\upsilon_\beta$ of the velocities $\upsilon_\beta\sim\partial_\beta\Phi$ across the protohalo and hence the tidal field $\Phi_{\alpha\beta}$
\begin{equation}
\Phi_{\alpha\beta}=\partial_\alpha\partial_\beta\Phi
\end{equation}
which is responsible for angular momentum generation. The mass distribution inside the protohalo itself is described by its inertia tensor $I_{\alpha\beta}$, 
\begin{equation}
I_{\alpha\beta} = \Omega_m\rho_\mathrm{crit}\:a^3\:\int_{V_L}\dd^3q\: (\vecq-\barq)_\alpha(\vecq-\barq)_\beta
\end{equation}
i.e. the second moments of the matter distribution, with the centre of mass at the position $\barq$ and the integration comprising the Lagrangian volume of the protohalo. Throughout, we use Einstein's summation convention.

The Levi-Civita-symbol in eqn.~(\ref{eqn_tidal_shearing}) generates the interesting misalignment property between the shear and inertia eigensystems which is required for generating angular momentum: Only the antisymmetric contribution $X^-_{\beta\gamma}$, derived from the commutator $X^-_{\beta\gamma} = [I_{\beta\gamma}, \Phi_{\beta\gamma}]$, to the product between the tensors $I_{\beta\gamma}$ and $\Phi_{\beta\gamma}$ is non-vanishing in contraction with the antisymmetric $\epsilon_{\alpha\beta\gamma}$ and therefore relevant for angular momentum generation. The symmetric contribution $X^+_{\beta\gamma}$, which can be isolated using the anticommutator $X^+_{\beta\gamma} = \{I_{\beta\gamma}, \Phi_{\beta\gamma}\}$ cancels in the contraction. This means that for angular momentum build-up, the tidal shear and the inertia tensors are not allowed to have a common eigensystem and be skewed relative to each other \citep{2009IJMPD..18..173S, 2012MNRAS.421.2751S}. Likewise, degeneracies in this relation evoked by spatial symmetries in the two tensors can prohibit the generation of angular momentum. Alternative models of galaxy angular momenta assume that the haloes are spun up by non-central, anisotropic infall in filaments \citep[see, for instance,][]{2010AIPC.1241.1108P, 2011arXiv1106.0538K, 2012arXiv1201.5794C}.

\subsection{Galaxy ellipticities}
Ellipticity correlations between galaxies are traced back to correlated angular momenta of their host haloes. CDM haloes acquire their angular momentum by tidal shearing and due to the fact that neighbouring galaxies experience correlated tidal fields, their angular momenta are correlated in consequence. The direction of the angular momentum $\vecl$ in turn determines the angle of inclination under which the galactic disk is viewed, and ultimately the ellipticity which is attributed to the galactic disk \citep{2000MNRAS.319..649H, 2001ApJ...559..552C, 2002ApJ...568...20C, 2002MNRAS.332..788M, 2003MNRAS.339..711H}: Linking the angular momentum direction $\lhat = \vecl/L$ to the components of the complex ellipticity $\epsilon$ using the above argument yields
\begin{equation}
\epsilon=\epsilon_+ +\ci\epsilon_\times
\quad\mathrm{with}\quad
\epsilon_+ = \alpha\frac{\hat{L}_x^2-\hat{L}_y^2}{1+\hat{L}_z^2},\quad
\epsilon_\times = 2\alpha\frac{\hat{L}_x\hat{L}_y}{1+\hat{L}_z^2},
\end{equation}
if the coordinate system is aligned with its $z$-axis being parallel to the line of sight. A rotation of the coordinate frame by $\varphi$ causes the complex ellipticity to rotate twice as fast, $\epsilon\rightarrow\exp(2\ci\varphi)\epsilon$. $\alpha$ is a free parameter weakening the dependence between inclination angle and ellipticity for thick galactic disks and has been determined to be $\alpha\simeq0.75$ in the APM sample \citep{2001ApJ...559..552C} with a large uncertainty.

It should be emphasised that the assumption of a galactic disk forming perpendicularly to the host halo angular momentum direction is a very strong one, which seems suggestive but has only little support from structure formation simulations. In fact, a number of studies point at possibly large misalignments and underline the complexity of the baryonic physics on galactic scales \citep{2002ApJ...576...21V, 2004ApJ...613L..41N, 2005ApJ...627L..17B, 2005ApJ...627..647B, 2008ASL.....1....7M, 2011arXiv1106.0538K}. In our analysis, misalignments between the symmetry axis of the galactic disk and the angular momentum axis of the host halo could be incorporated in choosing a smaller value for the disk thickness parameter $\alpha$, which will play the role of normalising the ellipticity spectra. The angular momentum-based alignment model is only able to capture the physics of tidal alignment of spiral galaxies. In the case of elliptical galaxies, a model which is linear in the tidal shear is more appropriate.

\subsection{Weak gravitational lensing}
The weak lensing convergence $\kappa$ provides a weighted line-of-sight measurement of the matter density $\delta$ \citep[for reviews, see][]{1999ARA&A..37..127M, 2001PhR...340..291B, 2002PhRvD..65f3001H, 2008ARNPS..58...99H, 2010CQGra..27w3001B}
\begin{equation}
\kappa = \int_0^{\chi_H}\dd\chi\: W_\kappa(\chi)\delta,
\end{equation}
with the weak lensing efficiency $W_\kappa(\chi)$ as the weighting function,
\begin{equation}
W_\kappa(\chi) = \frac{3\Omega_m}{2\chi_H^2}\frac{D_+}{a}G(\chi)\chi,
\mathrm{~with~}
G(\chi) = 
\int_\chi^{\chi_H}\dd\chi^\prime\:n(z)\frac{\dd z}{\dd\chi^\prime}\frac{\chi^\prime-\chi}{\chi^\prime}.
\end{equation}
$n(z)$ denotes the redshift distribution of the lensed background galaxies \citep[with the parameterisation introduced by][]{1995MNRAS.277....1S},
\begin{equation}
n(z) = n_0\left(\frac{z}{z_0}\right)^2\exp\left(-\left(\frac{z}{z_0}\right)^\beta\right)\dd z
\quad\mathrm{with}\quad \frac{1}{n_0}=\frac{z_0}{\beta}\Gamma\left(\frac{3}{\beta}\right).
\label{eqn_redshift}
\end{equation}
$z_0$ has been chosen to be $\simeq0.64$ such that the median of the redshift distribution is 0.9, which is anticipated for the EUCLID galaxy sample \citep{2007MNRAS.381.1018A, 2012arXiv1206.1225A}. With these definitions, one can carry out a Limber-projection \citep{1954ApJ...119..655L} of the weak lensing convergence for obtaining the angular convergence spectrum $C_\kappa(\ell)$, 
\begin{equation}
C_\kappa(\ell) = \int_0^{\chi_H}\frac{\dd\chi}{\chi^2}\:W_\kappa^2(\chi) P(k=\ell/\chi),
\end{equation}
which describes the fluctuation statistics of the convergence field. We will always work in the weak lensing regime, $\kappa,\gamma\ll1$, and approximate the reduced shear $g \equiv \gamma / (1-\kappa)$ with the lensing shear $\gamma$, which has the same statistical properties as the weak lensing convergence $\kappa$.

\section{Ellipticity correlations}
\label{sect_ellipticity}

\subsection{Angular momentum induced ellipticity correlations}
\label{sec:ecorr}
The idea behind intrinsic correlations is that neighbouring galaxies build up their angular momenta with correlated tidal shears because the galaxy separation is typically smaller than the correlation length of the tidal shear field. Under the assumption that the galactic disk orients itself perpendicular to the angular momentum direction of the host halo \citep[for a review on angular momenta of galactic disks, see][]{2012arXiv1207.4189R, 2012arXiv1207.4555B}, one perceives neighbouring galactic disks under correlated angles of inclination, and therefore the apparent shapes are correlated, which is measured in terms of ellipticities. We use two ellipticity correlation models in this paper, which are both constructed on the idea of correlated angular momenta, but which differ in their particular ansatz. The first model, proposed by \citet{2001ApJ...559..552C} establishes the link between the angular momentum direction to the tidal shear field in a random process in real space, whereas the second model, which is due to \citet{2002MNRAS.332..788M}, directly formulates the ellipticity field in Fourier space, which makes it easier to quantify ellipticity spectra, but whose parameterisation is not as clear as in the first case.

\subsubsection{ellipticity correlations}
Correlations of the two ellipticity components $\epsilon_+$ and $\epsilon_\times$ between two points $\vect_1$ and $\vect_2$ separated by an angular distance $\theta$ can be described using two correlation functions $C_{++}(\theta) = \bra \epsilon_+(\vect_1)\epsilon_+(\vect_2)\ket$ and $C_{\times\times}(\theta) = \bra\epsilon_\times(\vect_1)\epsilon_\times(\vect_2)\ket$, which are conveniently combined into two correlation functions $C_\pm(\theta)$,
\begin{eqnarray}
C_+(\theta) & = & 
C_{++}(\theta) + C_{\times\times}(\theta)\\
C_-(\theta) & = & 
C_{++}(\theta) - C_{\times\times}(\theta)
\end{eqnarray}
using $C_{+\times}(\theta) = \bra\epsilon_+(\vect_1)\epsilon_\times(\vect_2)\ket = 0$. Finally, ellipticity correlation functions can be transformed to the spectra $C_E^\epsilon(\ell)$ and $C_B^\epsilon(\ell)$ of the gradient and vorticity modes of the ellipticity field,
\begin{eqnarray}
C_E^\epsilon(\ell) & = & \pi\int\theta\dd\theta 
\left[C_+(\theta)J_0(\ell\theta) + C_-(\theta)J_4(\ell\theta)\right],
\label{eqn_e_transform}\\
C_B^\epsilon(\ell) & = & \pi\int\theta\dd\theta 
\left[C_+(\theta)J_0(\ell\theta) - C_-(\theta)J_4(\ell\theta)\right],
\label{eqn_b_transform}
\end{eqnarray}
by Fourier transform \citep{1992ApJ...388..272K, 2002A&A...389..729S, 2007A&A...462..841S,2010MNRAS.401.1264F}. Gravitational lensing in the lowest approximation is only able to excite $E$-modes in the ellipticity field.

\subsubsection{configuration space approach}
In this work we use the angular momentum-based ellipticity correlation model proposed by \citet{2001ApJ...559..552C} (referred to as the CNPT-model), who trace ellipticity correlations back to tidal shear correlations using the conditional probability distribution $p(\vecl|\Phi_{\alpha\beta})\dd\vecl$ introduced by \citet{2001ApJ...555..106L}: In this model, the distribution $p(\vecl|\Phi_{\alpha\beta})\dd\vecl$ is assumed as being Gaussian which is then being marginalised over the magnitude of the angular momentum vector, retaining only its directional dependence. Writing down the ellipticity components as a function of the angular momentum direction and employing the covariance $\bra L_\alpha L_\beta\ket$ as a function of the squared tidal shear tensor, as advocated by Lee and Pen, it is possible to relate the tidal shear correlations to the spectrum of the density field.

Angular momenta $\vecl$ are described as being coupled to the tidal shear by means of a Gaussian random process $p(\vecl|\Phi_{\alpha\beta})\dd\vecl$ involving tidal fields $\Phi_{\alpha\beta}$ shaping the covariance $\mathrm{cov}(L)_{\alpha\beta}$ of the Gaussian distribution \citep{2001ApJ...555..106L},
\begin{equation}
\mathrm{cov}(L)_{\alpha\beta} = \bra L_\alpha L_\beta\ket = \frac{\bra\vecl^2\ket}{3}\left(\frac{1+a}{3}\delta_{\alpha\beta} - a\: (\hat\Phi^2)_{\alpha\beta}\right),
\end{equation}
with the misalignment parameter $a$, which describes the average orientation of the protohalo's inertia to the tidal shear eigensystem. $a$ has been measured in numerical simulation to be close to 0.25 which we will assume in this work. $\hat{\Phi}$ is the unit normalised traceless tidal shear with the properties $\trace(\hat\Phi)=0$ and $\trace(\hat\Phi^2)=1$. This description is valid on scales where the correlations between inertia tensors are negligible.

The conditional probability density can be used for establishing a direct relation between ellipticity $\epsilon$ and tidal shear $\hat{\Phi}_{\alpha\beta}$ by integrating out angular momentum direction and magnitude:
\begin{equation}
\epsilon(\hat{\Phi}_{\alpha\beta}) = 
\int\dd\lhat\: \epsilon(\lhat)\int L^2\dd L\: p(\vecl|\Phi_{\alpha\beta})
\end{equation}
With this relation, one can write down the two correlation functions $\bra\epsilon_+(\bmath{x}_1)\epsilon_+(\bmath{x}_2)\ket$ and $\bra\epsilon_\times(\bmath{x}_1)\epsilon_\times(\bmath{x}_2)\ket$ of the three-dimensional ellipticity field in terms of moments $\zeta_n(r)$ \citep[see][]{2001ApJ...559..552C} of the tidal shear field. Those moments, in turn, are expressed as weighted integrals over the CDM-spectrum, where we impose a Gaussian smoothing on a scale of $10^{11}M_\odot/h$, which is typical for galaxies.

The correlation function of the 3-dimensional ellipticity field can then be projected onto the angular correlation function of the ellipticity components by using the configuration-space Limber-equation \citep{1954ApJ...119..655L}:
\begin{eqnarray}
C_{++}(\theta) & = & 
\int\dd\chi_1W_\epsilon(\chi_1)\:\int\dd\chi_2W_\epsilon(\chi_2)\:\bra\epsilon_+(\bmath{x}_1)\epsilon_+(\bmath{x}_2)\ket
\\
C_{\times\times}(\theta) & = &
\int\dd\chi_1W_\epsilon(\chi_1)\:\int\dd\chi_2W_\epsilon(\chi_2)\:\bra\epsilon_\times(\bmath{x}_1)\epsilon_\times(\bmath{x}_2)\ket
\label{eqn_real_limber}
\end{eqnarray}
with the distance distribution $W_\epsilon(\chi) = n(z(\chi))\dd z/\dd\chi$ resulting for a given cosmology from the observed redshift distribution $n(z)\dd z$ of background galaxies (see eqn.~\ref{eqn_redshift}). The separation distance entering the three dimensional correlation functions is completely determined by the two line-of-sight distances $\chi_1$, $\chi_2$ and the angle of separation $\theta$. \citet{2012arXiv1202.1196G} have shown that distorsions of the intrinsic ellipticity pattern due to the peculiar motion of galaxies is very small for multipoles up to $\ell=1000$.

\subsubsection{Fourier-approach}
We extend our analysis by the approach of \citet{2002MNRAS.332..788M} (abbreviated as MWK). Similar to \citet{2001ApJ...559..552C} they also work in the framework of tidal torque theory and relate intrinsic ellipticity to angular momentum assuming that the disk of a galaxy forms perpendicular to its spin axis. However, when computing the angular momentum the MWK-model entirely neglects any correlation between the tidal field and the tensor of inertia. They argue that due to the different correlation lengths involved (while the correlations for the inertia tensor primarily arise from smaller scales the correlations in the tidal field are long-ranged) a successive averaging-process is permissible. First, they perform an average over inertia tensors then over the tidal field expecting the ellipticity correlation arising from long-range correlations of the latter. In contrast to this, the CNPT-model takes explicitly the correlations of the inertia tensor and the tidal field via the misalignment parameter $a$ into account.

One more simplification made by MWK is to drop the dependence of the observed ellipticity on the $z$-component of the angular momentum of the galaxy, i.e.
\begin{equation}
\epsilon_+ = C \left(L_x^2-L_y^2 \right) 
\qquad \mathrm{and} \qquad 
\epsilon_{\times} = 2C\: L_xL_y.
\end{equation}
with a constant $C$. Accordingly, the intrinsic ellipticity scales quadratically with the modulus of the angular momentum, leaving faster spinning galaxies more flattened. Note that by the basic structure of this relation the symmetry properties of the ellipticity field are the same ones as in eqn.~(\ref{eqn_linear_alignment}).

Assuming shear and inertia being statistically independent allows to work completely in harmonic space, which greatly facilitates the computation. Since we aim at the power spectra of the intrinsic ellipticity it is convenient to introduce the parity conserving ($E$-mode) and parity violating ($B$-mode) part of the intrinsic ellipticity field
\begin{eqnarray}
  E(\mathbf k)\, k^2 &=& \left(k_x^2 - k_y^2 \right) \epsilon_+ (\mathbf k) + 2 k_xk_y \epsilon_\times (\mathbf k)\nonumber\\
  B(\mathbf k)\, k^2 &=& -2 k_xk_y \epsilon_+ (\mathbf k) + \left(k_x^2 -  k_y^2 \right) \epsilon_\times (\mathbf k).
\end{eqnarray}
Focusing on modes perpendicular to the line of sight one can derive the following dimensionless ellipticity power spectra for the $E$- and $B$-mode, respectively
\begin{eqnarray}
 \Delta_{X}^2(k) &=& \frac{C}{225} \left( \frac{3}{2}\Omega_m H_0^2 \right)^4 \int_0^\infty\frac{\dd \alpha}{\alpha} \Delta^2(\alpha k)\nonumber\\
 &  &\times \int_{-1}^1 \dd \mu \frac{\Delta^2 (k\sqrt{1+\alpha^2 - 2\alpha\mu})}{(1+\alpha^2-2\alpha\mu)^{7/2}}g_{X}(\alpha,\mu).
\end{eqnarray}
Here $X \in \{E, B\}$ and $g_{X}$ is a polynomial given in eqn.~(17) of \citet{2002MNRAS.332..788M} together with a detailed derivation of the expressions given above. As before, we smooth the linear power spectrum on an appropriate length scale with a Gaussian filter function.

Finally, in order to get the corresponding angular power spectra we have to again make use of the Fourier-space variant of Limber's projection \citep{1954ApJ...119..655L}
\begin{equation}
\frac{\ell (2\ell + 1)}{4\pi}\: C^\epsilon_{X}(\ell)= \frac{\pi}{\ell} \int_0^\infty \chi\,\dd \chi \, W^2_\epsilon(\chi) \Delta_{X}^2 (\ell / \chi)
\end{equation}
with the weighting function $W_\epsilon(\chi)\dd\chi = n(z)\dd z$ already introduced in eqn.~(\ref{eqn_redshift}).

\citet{2002MNRAS.332..788M} determine the constant $C$ by computing the expectation value of the squared angular momentum modulus and adjusting $C$ to match the mean-square source ellipticity typically observed in galaxy surveys. For our purpose, however, it makes more sense to choose $C$ in such a way that the angular power spectra obtained with the two different approaches coincide on largest scales. This is justified by the expectation that the large-scale power will be least effected by the differences in the two approaches under consideration. Hence, we first compute the spectra using formulae~(\ref{eqn_e_transform}) and~(\ref{eqn_b_transform}) for appropriately chosen misalignment parameter $a$ and galaxy thickness parameter $\alpha$ and subsequently determine $C$, so that the two models yield identical predictions for the variance of the intrinsic ellipticity field on large scales.

\subsubsection{comparison of the two ellipticity-models}
\begin{figure}
\begin{center}
\resizebox{\hsize}{!}{\includegraphics{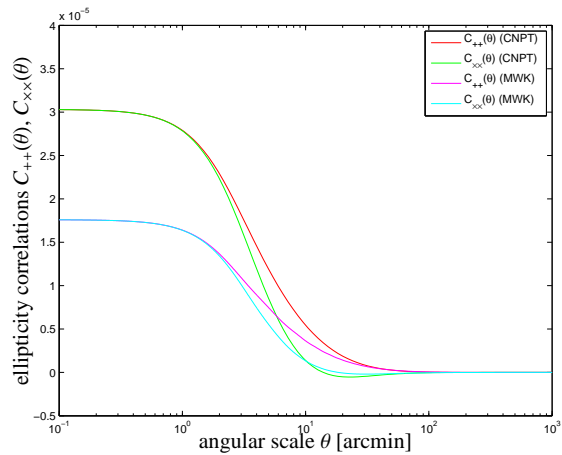}}
\end{center}
\caption{Angular ellipticity correlation functions $C_{++}(\theta)$ (green and magenta lines) and $C_{\times\times}(\theta)$ (red and cyan lines), for a smoothing scale of $M=10^{11}M_\odot/h$, a misalignment parameter $a=0.25$ and a disk thickness of $\alpha=0.75$. The correlation functions were derived using the CNPT- and MWK-models with the relative normalisation as discussed in the text.}
\label{fig_correlation}
\end{figure}

The ellipticity correlation functions $C_{++}(\theta)$ and $C_{\times\times}(\theta)$ resulting from both models are plotted in Fig.~\ref{fig_correlation} as a function of angle of separation $\theta$, where the projection was carried out for the EUCLID galaxy redshift distribution. The plot suggests correlation lengths of $\sim10$ arcminutes for the ellipticity field and shows that under the normalisations chosen, the correlation functions resulting from the CNPT-model achieves 50\% higher amplitudes in comparison to those predicted by the MWK-model, but otherwise the general shape is in very good agreement.

\begin{figure}
\begin{center}
\resizebox{\hsize}{!}{\includegraphics{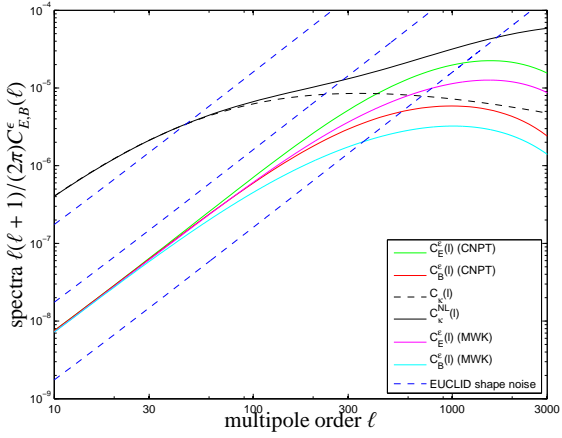}}
\end{center}
\caption{Ellipticity spectra $C_E^\epsilon(\ell)$ (green and magenta lines) and $C_B^\epsilon(\ell)$ (red and cyan lines) derived with the CNPT- and MWK-models, with weak convergence power spectrum $C_\kappa(l)$, both linear and nonlinear (black dashed and solid lines, respectively), and the EUCLID shape noise $\sigma_\epsilon^2/\bar{n}$, with $10^n\times\sigma_\epsilon^2/\bar{n}$, $n=0,1,2$ (blue dashed lines).}
\label{fig_spectrum}
\end{figure}

Computing the spectra $C_E^\epsilon(\ell)$ and $C_B^\epsilon(\ell)$ yields Fig.~\ref{fig_spectrum}, where for comparison the linear and nonlinear spectra $C_\kappa(\ell)$ for the weak lensing convergence and the EUCLID shape noise levels $\sigma_\epsilon^2/n$ are plotted. The shape of the ellipticity spectra shows constant amplitudes up to scales of $\ell\sim300$, where individual ellipticities are uncorrelated, and correlations between ellipticities are present on smaller angular scales. The spectra exhibit a wide maximum on multipoles of $\ell\sim10^3$ before dropping in amplitude, which is caused by imposing the mass cutoff. For comparison and motivating our analysis we plot predictions for the weak lensing spectrum $C_\kappa(\ell)$ for linear and nonlinear CDM spectra, as well as the anticipated shot-noise for EUCLID. Clearly, intrinsic ellipticity correlations are subdominant compared to weak lensing induced ellipticity correlations, but can in amplitude amount to up to 30\% of the lensing signal on multipoles of $\ell\simeq10^3$ before the shape noise makes measurements difficult. Comparing the two ellipticity models show that, if the normalisation is chosen as explained, the MWK-model predicts lower spectra than the model by CNPT by about 50\% on high multipoles as in the case of the correlation function at small separations, but both models predict similar ratios between $C^\epsilon_E(\ell)$ and $C^\epsilon_B(\ell)$ amounting to about a factor of 5 at high multipoles. Interestingly, intrinsic ellipticity correlations would dominate over the weak lensing signal if the lensing prediction was derived using linear structure growth only. Comparing the spectra $C^\epsilon_E(\ell)$ and $C_\kappa(\ell)$ with the shape noise levels of EUCLID clearly demonstrate the importance of intrinsic ellipticity correlations in weak lensing data. 

The $B$-mode spectrum $C^\epsilon_B(\ell)$, which is sourced by intrinsic alignments, is smaller by more than one order of magnitude compared to the $E$-mode spectrum $C^\epsilon_E(\ell)$ at high multipoles but might dominate over other higher-order lensing effects which are able to excite parity-violating modes in the ellipticity field such as source-lens clustering \citep{2002A&A...389..729S}, multiple lensing along the line-of-sight or violations of the Born-approximation \citep{2002ApJ...574...19C,2006JCAP...03..007S, 2012MNRAS.420..455S}.

\subsection{Variance in apertures}\label{ssec:msap}
Quantities derived from the weak lensing spectrum $C_\kappa(\ell)$ are weighted variances of the convergence inside apertures of varying size $\theta$, introduced by \citet{1996MNRAS.283..837S}:
\begin{equation}
\bra \kappa^2 \ket(\theta) = \frac{2}{\pi}\int\ell\dd\ell\: W^2_1(\ell\theta)\:C_{\kappa}(\ell),
\end{equation}
and
\begin{equation}
 \bra M_{\kappa}^2 \ket(\theta) = \frac{2}{\pi}\int\ell\dd\ell\:W^2_4(\ell\theta)\:C_{\kappa}(\ell),
\end{equation}
which measure the scale-dependence of fluctuations' variance in the respective field. The weighting functions $W_0(x)$, and $W_4(x)$, $x=\ell\theta$, are defined as:
\begin{equation}
W_1(x) = \frac{J_1(x)}{x}
\quad\mathrm{and}\quad
W_4(x) = \frac{12J_4(x)}{x^2},
\end{equation}
respectively, for the shear variance averaged in an aperture of size $\theta$ and the aperture mass variance. The weak lensing power spectrum is substituted in the previous definitions with $C_E^\epsilon(\ell)$, and $C_B^\epsilon(\ell)$ in order to obtain:
\begin{eqnarray}
\bra \epsilon^2_+ \ket(\theta) & = & \frac{2}{\pi} \int\ell\dd\ell\:W^2_1(\ell\theta)C_E^\epsilon(\ell),\\
\bra M_\mathrm{ap}^2 \ket(\theta) & = & \frac{2}{\pi} \int\ell\dd\ell\: W^2_4(\ell\theta)C_E^\epsilon(\ell),\\
\bra \epsilon^2_\times \ket(\theta) & = & \frac{2}{\pi}\int\ell\dd\ell\: W^2_1(\ell\theta)C_B^\epsilon(\ell),\\
\bra M_{\perp}^2 \ket(\theta) & = & \frac{2}{\pi} \int\ell\dd\ell\: W^2_4(\ell\theta)C_B^\epsilon(\ell),
\end{eqnarray}
which are analogous quanitites if the origin of ellipticity correlations would be purely intrinsic and generated by correlated angular momenta.

\begin{figure}
\begin{center}
\resizebox{\hsize}{!}{\includegraphics{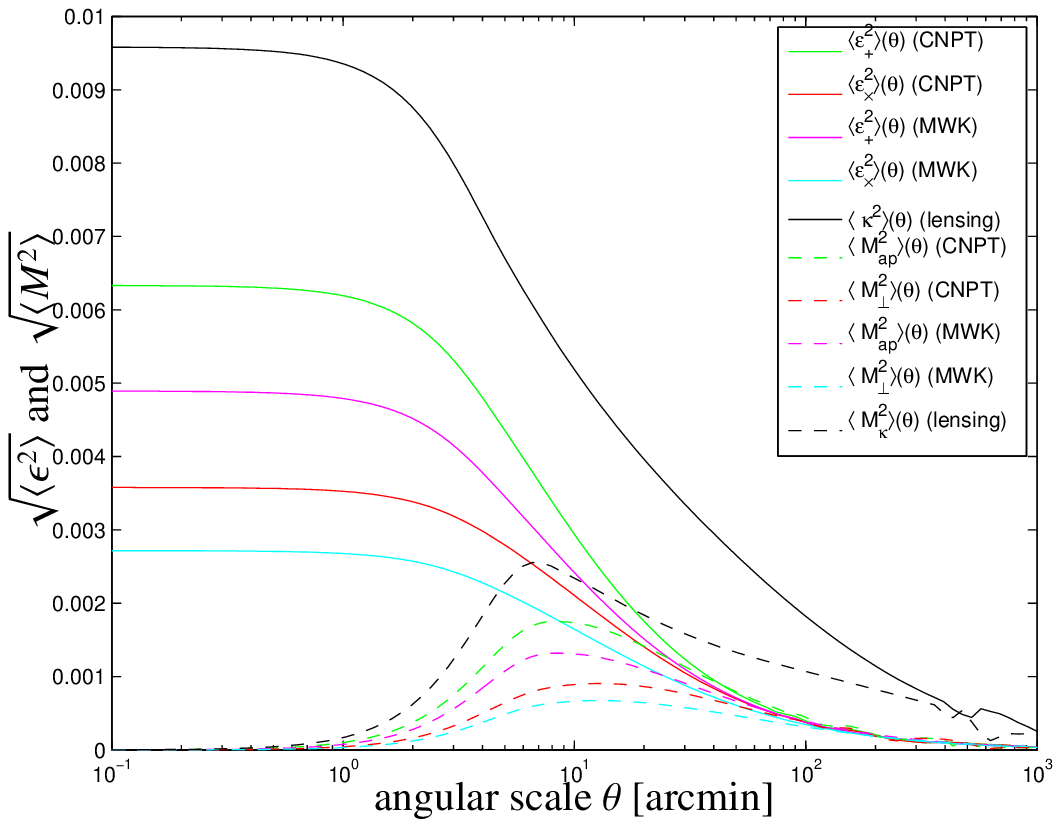}}
\end{center}
\caption{Standard deviation of the averaged ellipticity and aperture mass as a function of aperture size $\theta$ for the CNPT- and MWK-models: tangential ellipticity $\bra\epsilon_+^2\ket$ (solid green and magenta lines), radial ellipticity $\bra\epsilon_\times^2\ket$ (solid red and cyan lines), aperture mass $\bra M_\mathrm{ap}^2\ket$ (dashed green and magenta lines) along with $\bra M_\perp^2\ket$ (dashed red and cyan lines), all for the EUCLID galaxy sample in comparison to the same quantities derived from the weak lensing convergence (corresponding black lines).}
\label{fig_variance}
\end{figure}

In Fig.~\ref{fig_variance} we show, how the aperture-weighted variances of the intrinsic ellipticity field should behave in the EUCLID galaxy sample as a function of angular scale in comparison to that of the weak lensing field. We consider both tangential and radial shears and compare the results between the two different intrinsic ellipiticity models.  

The aperture-weighted variances derived from intrinsic ellipticities decrease in magnitude which of course is a generally expected behaviour caused by the weighting functions $W_n(\ell\theta)$, and exhibit lower amplitudes compared to the lensing ones, as can be expected from the relative magnitudes of the spectra. From angular scales of 100 arcminutes on intrinsic alignments have dropped to zero, which is compatible with them being a small-scale phenomenon, while weak lensing still has a considerable signal. As expected, the differnece in magnitude of the variances sourced by $E$-modes and $B$-modes is smaller than the difference in spectra on small scales because of the averaging in multipole $\ell$, and an analogous argument applies to the predictions by the two models under consideration. 

\section{Parameter likelihood}
\label{sect_likelihood}
In this section, the dependence of the intrinsic ellipticity spectrum on the cosmological parameter set is investigated. This is of particular relevance because of their contaminating effect in weak lensing data by introducing spurious ellipticity correlations, and because they depend on the cosmological model in a very nonlinear way, much stronger than e.g. the weak lensing convergence: We point out that in our models, the angular momentum $\vecl$ reflects the squared tidal shears $\partial_i\partial_j\Phi$, and the ellipticity field $\epsilon$ in turn has a very complex dependence on the angular momentum direction, which can be approximated to be quadratic for small line of sight-components of the angular momentum direction. In the following, we keep the parameter $\alpha$ in our ellipticity model constant, because it can in principle be determined by analysing morphological data, as \citet{2001ApJ...559..552C} demonstrated. The misalignment parameter $a$ replaces $\sigma_8$ because it fixes the normalisation of the spectra, and is completely degenerate with $\alpha$.

\subsection{Parameter dependences of intrinsic alignments}
When considering the angular spectra describing the ellipticity field, the parameter sensitivity of intrinsic alignments can be nicely illustrated by considering derivatives of the spectra with respect to cosmological parameters, weighted by the inverse noise. For illustration, we assume that weak lensing-induced ellipticity alignments were absent from the data, and that the ellipticity shape noise would be that of EUCLID. Therefore the sensitivities, i.e. the derivative of the observables with respect to the parameters to be estimated in units of the noise,
\begin{equation}
\frac{1}{\sqrt{\mathrm{cov}_{X}(\ell)}}\frac{\partial C^\epsilon_X}{\partial x_\mu},\quad X\in\left\{E,B\right\},
\end{equation}
correspond to the contributions $\sqrt{\dd F_{\mu\mu}/\dd\ell}$ to the diagonal entries of a Fisher-matrix $F_{\mu\nu}$, which describes the parameter dependence of the spectra $C^\epsilon_E(\ell)$ and $C^\epsilon_B(\ell)$ on a cosmological model. The covariances of acquire a cosmic variance error and a Poissonian shape measurement error,
\begin{equation}
\mathrm{cov}_X(\ell) = \frac{2}{2\ell+1}\frac{1}{f_\mathrm{sky}}\left(C^\epsilon_X(\ell) + \frac{\sigma^2_\epsilon}{n}\right)^2
\end{equation}
with $\sigma_\epsilon = 0.3$, $n = 30/\mathrm{arcmin}^2$ as the number density of galaxies per square steradian and the sky fraction $f_\mathrm{sky}=1/2$.

\begin{figure}
\begin{center}
\resizebox{\hsize}{!}{\includegraphics{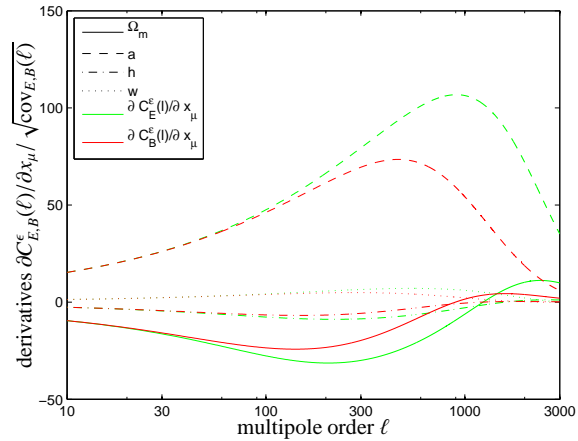}}
\end{center}
\caption{Sensitivities $\partial C_E^\epsilon(\ell)/\partial x_\mu$ (green lines) and $\partial C_B^\epsilon(\ell)/\partial x_\mu$ (red lines) in units of $\sqrt{\mathrm{cov}_X(\ell)}$ as a function of maximum multipole order $\ell$, with respect to the cosmological parameters $\Omega_m$ (dashed lines), $a$ (solid lines), $h$ (dash-dotted lines) and $w$ (dotted lines). The spectra were computed with the CNPT-model.}
\label{fig_derivative}
\end{figure}

We depict these quantities in Fig.~\ref{fig_derivative} for the basic set of cosmological parameters considered here: $x_\mu\in\left\{\Omega_m,a,h,w\right\}$, where $n_s$ has been omitted due to its very weak influence on the spectra. Clearly, the $E$-mode and $B$-mode spectra exhibit an identical behaviour on large, cosmic variance dominated scales, where they reflect identical dependence on the physical processes of angular momentum generation and disk orientation, before differening on multipoles $\ell\gsim300$, where weaker $B$-modes start being influenced by the noise level. The effectively non-existent dependence of the ellipticity spectra on the dark energy equation of state parameter $w$ is particularly interesting and suggests that intrinsic alignments contaminations in weak lensing can be investigated almost independently from the dark energy model assumed. Conversely, the dependences on $\Omega_m$ and $a$ are particularly strong, because $a$ determines the amplitude of the spectra in much the same way as $\sigma_8$ fixes the normalisation of the weak lensing spectra.

\subsection{non-Gaussian likelihoods} \label{nongauss}
This section is intended to check whether our assumption of Gaussianity for the parameters likelihood is well-grounded. In order to achieve this we compare the Gaussian likelihoods derived by using the Fisher-formalism,
\begin{equation}
\mathcal{L}\propto\exp\left(-\frac{(x_\mu-x_\mu^\mathrm{fid})^2}{2\sigma_\mu^2}\right)
\quad\mathrm{with}\quad
\sigma_\mu^2 = \frac{1}{F_{\mu\mu}}
\end{equation}
where the Fisher-matrix is determined from the curvature of the logarithmic likelihood,
\begin{equation}
F_{\mu\nu} = 
\sum_{\ell = \ell_\mathrm{min}}^{\ell_\mathrm{max}}
\frac{\partial C_E^\epsilon(\ell)}{\partial x_\mu} \frac{1}{\mathrm{cov}_E(\ell)}\frac{\partial C_E^\epsilon(\ell)}{\partial x_\nu}
\end{equation}
with a direct evaluation of the respective likelihood function, derived using the relation:
\begin{equation}\label{likelihood}
\mathcal{L}\propto\exp\left(-\chi^2(x_\mu)/2\right)
\end{equation}
where the $\chi^2$-functional is given by
\begin{equation}\label{chi2}
\chi^2(x_\mu) = 
\sum_{\ell = \ell_\mathrm{min}}^{\ell_\mathrm{max}}
\frac{1}{\mathrm{cov}_E(\ell)}
\left[C_E^\epsilon(\ell|x_\mu) - C_E^\epsilon(\ell|x_\mu^\mathrm{fid})\right]^2,
\label{eqn_chi2}
\end{equation}
and quantifies the goodness-of-fit of a parameter choice $x_{\mu}$ in the space spanned by the set of cosmological parameters $\{\Omega_m,a,h,w,n_s\}$. It is important to specify that due to the weakness of the intrinsic ellipticity correlations we consider only conditional errors, i.e. we let one parameter vary at a time, keeping the remaining fixed and focus only on the stronger $E$-mode spectrum $C^\epsilon_E(\ell)$. The reduced dimensionality allows us to compute the likelihood directly on a grid without having to make use of Monte-Carlo sampling techniques for evaluating the likelihood.

Any deviation from a Gaussian shape of the likelihood $\mathcal{L}$ is caused by a nonlinear dependence of the spectrum $C^\epsilon_E(\ell)$ on a model parameter $x_\mu$, which is due to the fact that the $\chi^2$-functional deviates from a parabolic shape if the model parameter is varied. The likelihood assumes an approximately Gaussian shape if it is sufficiently peaked such that a Taylor-expansion of the nonlinear parameter dependences is applicable in the region around the fiducial parameter choice. In our case, non-Gaussian shapes have been observed if the summation in eqn.~(\ref{eqn_chi2}) was restricted to the multipole range $10\leq\ell\leq100$ and quickly became Gaussian if the summation was carried out to higher multipoles.

\begin{figure}
\resizebox{\hsize}{!}{\includegraphics{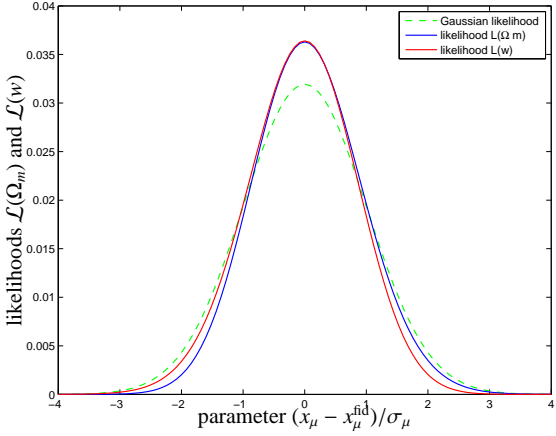}}
\caption{Conditional likelihoods $\mathcal{L}(\Omega_m)$ (solid blue line) and $\mathcal{L}(w)$ (solid red line), along with their Gaussian approximations derived with the Fisher-formalism ($\sigma_\mu = 1/\sqrt{F_{\mu\mu}}$, dashed green line), for an observation of ellipticity spectrum $C_E^\epsilon(\ell)$ with the EUCLID survey characteristics with all other parameters fixed to their fiducial values. The multipole range was set to $10\leq\ell\leq100$ and the spectrum $C^\epsilon_E(\ell)$ entering the likelihood-calculation resulted from the CNPT-model.}
\label{fig_pdf}
\end{figure}

The non-Gaussian likelihoods $\mathcal{L}(\Omega_m)$ and $\mathcal{L}(w)$ for the matter density and the dark energy equation of state can be seen in Fig.~\ref{fig_pdf} in comparison to their Gaussian approximation. All likelihoods are centered on the fiducial model value $x_\mu^\mathrm{fid}$ and scaled with the width $\sigma_\mu$ derived with the Fisher formalism. Most notably, the likelihoods are more strongly peaked than their Gaussian counterparts, with slight asymmetries of $\mathcal{L}(\Omega_m)$ towards large values and of $\mathcal{L}(w)$ towards small parameter values. It should be noted, that the misalignment parameter $a$ (together with the disk thickness $\alpha$) is a linear parameter in our models and its likelihood $\mathcal{L}(a)$ is always of Gaussian shape. We conclude that the amount of deviation from the ideal shape is not a serious impediment for applying the Fisher-formalism for investigating intrinsic alignments, keeping in mind that in reality one observes alignments over a much wider multipole range such that the likelihoods are closer to Gaussianity.

\begin{figure*}
\resizebox{\hsize}{!}{\includegraphics{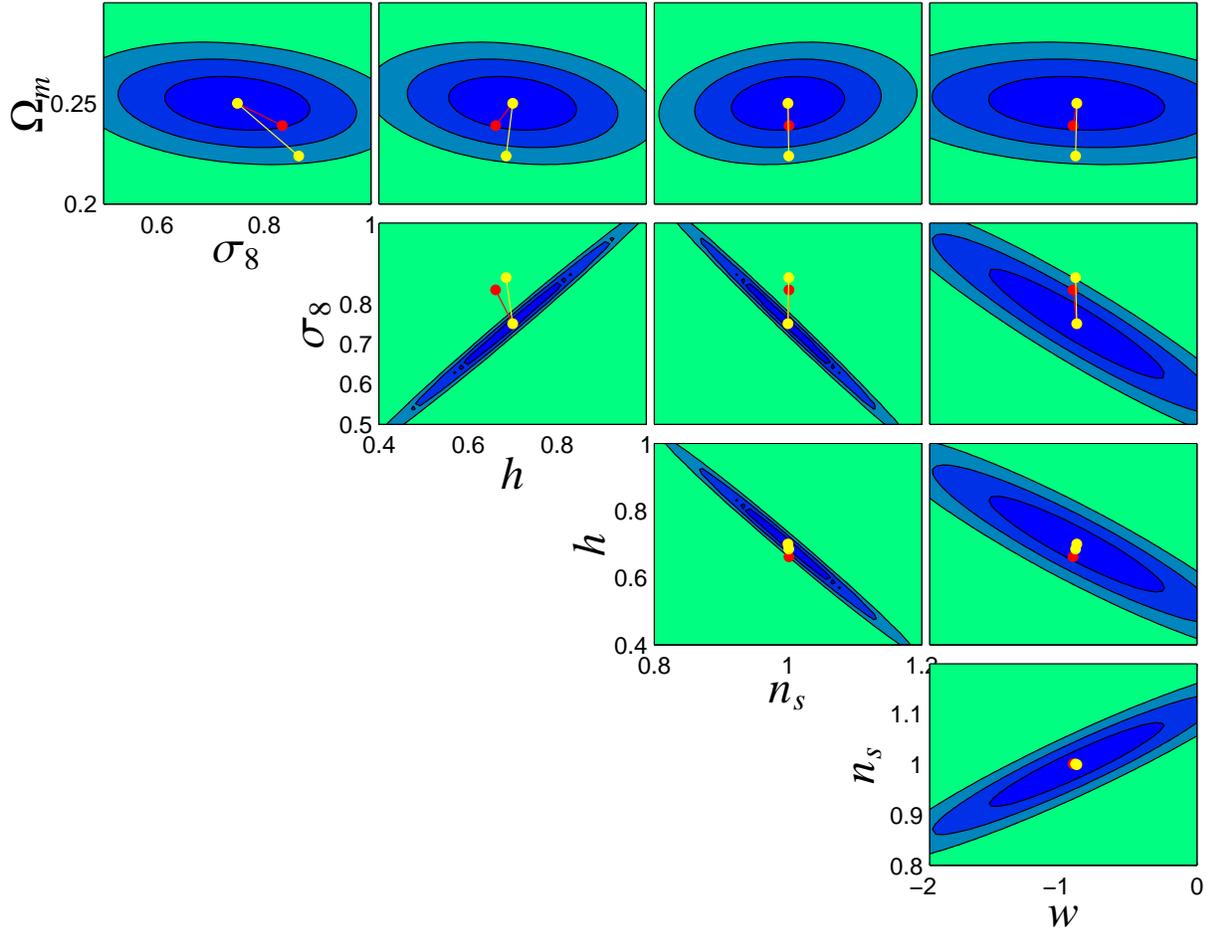}}
\caption{Parameter estimation biases in $\Omega_m$, $\sigma_8$, $h$, $n_s$ and $w$ obtained respectively with the CNPT-model (in red) and with the MWK-model (in yellow) for EUCLID's observation of the weak lensing spectrum $C_\kappa(\ell)$, which is contaminated by intrinsic alignments $C_E^\epsilon(\ell)$ on small scales. The misalignment parameter is set to $a = 0.25$; the disk thickness parameter is set to $\alpha=0.75$, the multipole range was $10\leq\ell\leq3000$ and as a noise amplitude we considered $\sigma_\epsilon=0.3$. The ellipses give $1,2,3\sigma$ statistical uncertainties on the cosmological parameters from the weak lensing spectrum $C_\kappa(\ell)$ from the same multipole range.}

\label{fig_bias}
\end{figure*}

\section{Interference with weak lensing} \label{sect_lensing}
In this section we show how the intrinsic alignments can affect measurements of the convergence spectrum by quantifying the parameter biases arising when trying to explain the data consisting of both weak lensing-induced ellipticity correlations and intrinsic alignments by a model that only accounts for weak lensing and neglects intrinsic alignments. We consider the case of EUCLID's weak lensing survey in a non-tomographic setup and give an estimation of the biases, if intrinsic alignments are not removed from data \citep[as proposed by][]{2005A&A...441...47K, 2009A&A...507..105J,2008A&A...488..829J} or not properly modelled \citep{2003A&A...398...23K,2002A&A...396..411K}. We aim to supplement previous analysis of intrinsic alignment contaminations such as \citet{2004PhRvD..70f3526H}, \citet{2007NJPh....9..444B} and \citet{2008MNRAS.389..173K} by using a physically motivated and well described alignment model for spiral galaxies with a small number of parameters which can be accessed by morphological galaxy samples (the disk thickness parameter $\alpha$) and cosmological simulations (the misalignment parameter $a$).

\subsection{Parameter constraints}\label{ssec:fisherr}
Statistical errors on constraints on the cosmological parameters from the projected weak lensing power spectrum $C_{\kappa}(l)$ when ignoring the intrinsic ellipticity spectrum can be easily obtained by using the Fisher matrix formalism, where the Fisher-matrix $F_{\mu\nu}$ measures the curvature of the logarithmic likelihood $\ln\mathcal{L}$ in all parameter directions \citep{1997ApJ...480...22T}:
\begin{equation} \label{fmunu}
F_{\mu\nu} = \sum_{\ell = \ell_\mathrm{min}}^{\ell_\mathrm{max}} \frac{\partial C_{\kappa}(\ell)}{\partial x_{\mu}} \frac{1}{\mathrm{cov}_\kappa(\ell)} \frac{\partial C_{\kappa}(\ell)}{\partial x_{\nu}}
\end{equation}
where the covariance is given by:
\begin{equation} \label{cov}
\mathrm{cov}_\kappa(\ell) = \frac{2}{2\ell+1}\frac{1}{f_\mathrm{sky}}\left(C_\kappa(\ell) + \frac{\sigma^2_{\epsilon}}{n}\right)^2.
\end{equation}
In the latter expression EUCLID's noise $\sigma^2_{\epsilon}/n$ was used as well as $f_\mathrm{sky}=1/2$ for the sky fraction, and the parameter space is spanned by the cosmological parameters $x_{\mu} \in \left\{ \Omega_m, \sigma_8, h, n_s, w \right\}$. We will use the covariance $\mathrm{cov}_\kappa(\ell)$ throughout this chapter and neglect small contributions due to intrinsic alignments to the covariance of the ellipticity field, as the covariance and therefore the sampling noise is dominated by weak lensing.


The statistical $n\sigma$-ellipses obtained as cross-sections through the Gaussian-approximated likelihood for all pairs of parameters are shown in Fig.~\ref{fig_bias} together with the systematical errors in parameter estimation if intrinsic alignments are not taken care of. The extend to which weak lensing parameter likelihoods are Gaussian is investigated in detail by \citet{2012arXiv1205.3984W}.

\subsection{Parameter estimation biases}\label{sect_bias}
Ultimately our analysis aims to quantify how biased the parameter estimation with EUCLID-data will be if intrinsic alignments as predicted from angular momentum models were present in the data but if we were to interpret the data with a model which does not take intrinsic ellipticities into account. We therefore identify a true model which includes intrinsic alignments
\begin{equation}
C_t(\ell) = C_{\kappa}(\ell) + C^{\epsilon}_E(\ell) + \frac{\sigma^2_{\epsilon}}{n}
\end{equation}
and a false model, which omits intrinsic alignments and considers the ellipticities as random,
\begin{equation}
C_f(\ell)= C_{\kappa}(\ell) + \frac{\sigma^2_{\epsilon}}{n}.
\end{equation}
For both models, one can specify a goodness-of-fit parameter which in the case of Gaussian errors is the $\chi^2$-functional. If the data, which follows the model $C_t(\ell)$, is interpreted with the wrong model $C_f(\ell)$, the corresponding $\chi^2$-functional will exhibit its minimum at a position in parameter space shifted from the true parameter choice, because the incomplete model is forced to provide a fit to the data by detuning the parameter set away from the fiducial values.
The way to achieve this goal was proposed by a number of authors \citep{2007MNRAS.381.1347C, 2008MNRAS.391..228A, 2009MNRAS.392.1153T, 2011MNRAS.tmp..612M, 2012MNRAS.tmp.3168S} in different contexts: Taking the second-order Taylor expansion of the wrong $\chi^2_f$-functional around the best-fit point $\boldmath{x}_t$ of the true model one retrieves an expression involving the vector $\bmath{\delta}$:
\begin{equation}
\chi^2_f(\bmath{x}_f) = \chi^2_f(\bmath{x}_t) + \sum_{\mu}\frac{\partial}{\partial x_{\mu}}\chi^2_f(\bmath{x}_t)\delta_{\mu} + \frac{1}{2} \sum_{\mu,\nu}\frac{\partial^2}{\partial x_{\mu}\partial  x_{\nu}}\chi^2_f(\bmath{x}_t)\delta_{\mu}\delta_{\nu},
\end{equation}
being $\bmath{\delta} \equiv \bmath{x}_f - \bmath{x}_t$. For weak systematics, this approach has been demonstrated to yield very accurate results for the estimation biases by comparison with the shift of the likelihood peak evaluated by MCMC-techniques \citep{2010MNRAS.404.1197T}.

Now, by extremising the ensamble-averaged $\langle \chi^2_f(\bmath{x}_f) \rangle$ one gets the best-fit position $\bmath{x}_f$. This operation yields a linear system of equations:
\begin{equation}
\sum_{\nu} G_{\mu \nu} \delta_{\nu} = a_{\mu}
\quad\rightarrow\quad
\delta_\mu = \sum_\nu (G^{-1})_{\mu\nu}a_\nu
\label{eqn_linear_solve}
\end{equation}
and can be inverted directly leading to the estimation bias $\bmath{\delta}$.
%
Substitution gives expressions for the quantities $G_{\mu \nu}$ and $a_{\mu}$ which involve derivatives of the spectra,
\begin{eqnarray}
G_{\mu \nu} &\equiv & 
\sum_{\ell = \ell_\mathrm{min}}^{\ell_\mathrm{max}}\frac{1}{\mathrm{cov}_{\kappa}}
\left[ \frac{\partial C_{\kappa}(\ell)}{\partial x_{\mu}}\frac{\partial C_{\kappa}(\ell)}{\partial x_{\nu}} - C_E^\epsilon(\ell)\frac{\partial^2 C_{\kappa}(\ell)}{\partial x_{\mu}\partial  x_{\nu}} \right], \nonumber \\
a_{\mu} &\equiv& 
\sum_{\ell = \ell_\mathrm{min}}^{\ell_\mathrm{max}}\frac{1}{\mathrm{cov}_{\kappa}}
\left[ C_E^\epsilon(\ell)\frac{\partial C_{\kappa}(\ell)}{\partial x_{\mu}} \right].
\end{eqnarray}
It is worth to notice that the espression for $G_{\mu \nu}$ simplifies to $F_{\mu\nu}$ when the correct model is used, and the bias vector is therefore zero. A consequence of this argument is that the inclusion of a Gaussian prior $F^\mathrm{prior}_{\mu\nu}$ would reduce the parameter estimation bias due to the transformation $G_{\mu\nu}\rightarrow G_{\mu\nu} + F^\mathrm{prior}_{\mu\nu}$, leading to smaller values for $\delta_\mu$ in the inversion of the linear system eqn.~(\ref{eqn_linear_solve}).

Fig.~\ref{fig_bias} shows the biases in the estimation of the cosmological parameters induced by considering the intrinsic ellipticities. We computed the $1,2,$ and $3\sigma$ ellipses by means of the Fisher matrix, as explained in Sect.~\ref{ssec:fisherr}, and we considered the full range of multipoles going from $\ell_\mathrm{min} = 10$ to $\ell_\mathrm{max}=3000$ well into the noise-dominated regime. We calculated the biases for both the CNPT and the MWK models of the intrinsic alignments: In what concerns the CNPT-model we considered the value of the misalignment parameter found in $n$-body simulations \citep{2000ApJ...532L...5L} $a=0.25$, and $\alpha = 0.75$ for the thickness of the disk. The biases are shown respectively in red for the CNPT-model, and in yellow for the MWK-model. All parameters apart from the dark energy equation of state $w$ and the slope $n_s$ are biased significantly by at least $1\sigma$, and in almost all cases the shift is not along the primary statistical degeneracy. The differences in the biases between the CNPT- and MWK-models reflect the difference in amplitude they predict, which is in our case due to the choice of normalisation.

The first thing to notice is how the biases depend on the parameters. Evidently $\Omega_m$ and $\sigma_8$, parameters on which the convergence spectrum $C_{\kappa}(\ell)$ highly depends, seem to be mostly affected, and this is in line with the fact that, in a Bayesian analysis, the fit will tweak preferably those parameters the model is more strongly dependent on, therefore generating larger biases. The estimation biases in $\Omega_m$ and $\sigma_8$ suggest that the presence of $C^\epsilon_E(\ell)$ is increasing the normalisation and adds power to the high-$\ell$ part of the spectrum in exactly the way which was expected by choosing higher $\sigma_8$-values and lower $\Omega_m$-values. For these two cosmological parameters, however, excellent priors from the CMB temperature and polarisation spectra are available which can be used for further reducing the biases.

Nonetheless our finding of substantial invariance of the dark energy parameter $w$ appears to be in contradiction with the results obtained by \citet{2012MNRAS.tmp.3339K}, who performed a joint analysis on galaxy shape and galaxy number density correlations in order to set constraints on cosmological parameters, thereby accounting for both intrinsic alignments and galaxy bias and found very large biases in $w$. It seems very interesting that the linear alignment model used by \citep{2012MNRAS.tmp.3339K} and our quadratic alignment model lead to such diverse conclusions. Due to the complication and difficulties of physically modelling the intrinsic alignments on the one hand, and due to the little information known about the galaxy bias on the other, in this work the two contaminants have been parametrized as functions of scale and redshift \citep[see also Sect.~2 of ][]{2007NJPh....9..444B} for further details on intrinsic alignment modelling), and their values at different $(k,z)$ points have been considered as nuisance parameters along with the cosmological parameters. The constraints on the cosmological parameters have been then obtained by marginalizing over the nuisance parameters, and the analysis of \citet{2010A&A...523A...1J} and \citet{2012MNRAS.tmp.3339K} ultimately show that the parameters which are mostly affected by the marginalization are the ones entering the dark energy equation of state $w_0$ and $w_a$. This illustrates the range of predictions from different alignment models and emphasises the need of a better theoretical understanding of galaxy alignments.

In fact, this was our motivation for investigating the problem of intrinsic alignments by adopting a physical model, although simplified, which allows to keep the statistics separated from systematics, and therefore to quantify the net contribution of the latter to the first. On the other hand, the choice of using larger sets of data and incorporating the biases by means of a parametrisation certainly increases the statistics, but necessarily leads to a mixing of statistics and systematics, since the biases are disguised as nuisance parameters to be marginalized over. This leads to internally calibrated constraints on the cosmological parameters, but makes it difficult to estimate to which extent the modelling of the data is actually affected by the inclusion of contaminating effects such as intrinsic alignments.

It is worth noting that for linearly evolving Gaussian random fields the quadratic alignment model does not yield a nonzero prediction for cross-correlations between intrinsic alignments and weak gravitational lensing (GI-alignments), as this correlation would be proportional to the third moment of a Gaussian random field. Only in the case of nonlinearly evolving density fields this correlation would be nonzero due to higher-order corrections in the weak lensing signal. But conversely, positive detections of GI-alignments on linear scales would be a clear signature of linear alignment models rather than quadratic ones and could help to choose the correct family of alignment models.

\subsection{Scaling of the estimation bias}
It is necessary to investigate how the estimation bias scales with the normalisation of the intrinsic alignment spectra, as the parameters $a$ and $\alpha$ in the CNPT-model and the resulting normalisation $C$ in the MWK-model have a large uncertainty. It is worth recalling that the misalignment parameter $a$ is measured in $n$-body simulations of structure formation, and the galaxy disk thickness $\alpha$ is taken from data on galaxy morphologies.

\begin{figure}
\resizebox{\hsize}{!}{\includegraphics{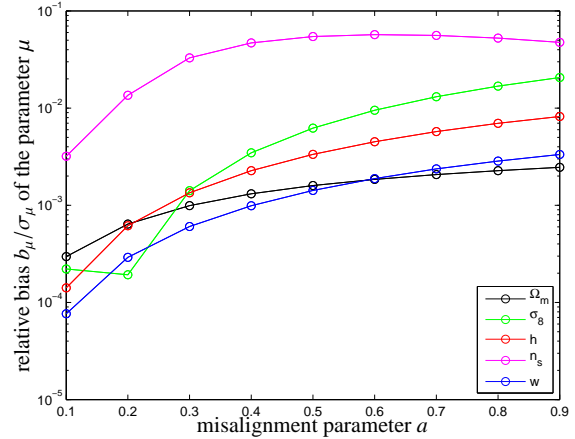}}
\caption{Biases $b_{\mu}$ in units of the statistical error $\sigma_{\mu}$ for the cosmological parameters $\Omega_m$ (black line), $\sigma_8$ (green line), $h$ (red line), $n_s$ (magenta line) and $w$ (blue line), as a function of the misalignment parameter $a$ of the CNPT-model.}
\label{fig_bias_a}
\end{figure}

In Fig.~\ref{fig_bias_a} we focus on the CNPT-model, by plotting the biases in units of the conditional error $\sigma_{\mu}^2 = 1/F_{\mu\mu}$ as a function of the misalignment parameter $a$ while keeping $\alpha$ fixed at 0.75. An increasing value of $a$ means a higher misalignment between shear and inertia tensors \citep{2000ApJ...532L...5L,2001ApJ...559..552C,2009IJMPD..18..173S}, and hence a higher correlation between angular momenta, or, differently phrased, less randomness in their directions. This means that, within the model, the angular momenta trace the underlying tidal shear field in a tighter way. The lower limit $a=0$ indicates therefore complete randomness and absence of any link of the angular momenta to the gravitational potential, as described in \citet{2000ApJ...532L...5L}. It is visible how the biases grow fast for greater values of the parameter $a$, simply meaning that stronger intrinsic correlations would represent a stronger contamination to the covergence spectrum, but that for very large values of $a$ the dependence of the parameter estimation bias with $a$ is saturated and it evolves weaker with increasing $a$, and even decreases in the case of the parameter $n_s$. Given the fact that the estimation biases $\delta_\mu/\sigma_\mu$ change by almost two decades as $a$ is varied, it is vitally important to determine $a$ beforehand, either from independent observations or from simulations of structure formation.

\subsection{Observations of intrinsic alignments}
Another approach to gain infomation about the intrinsic alignments relies on the subtraction of the weak lensing signal from the overall signal that is measured. This is, roughly speaking, the inverse of what is usually done with nulling techniques \citep[for details see][]{2008A&A...488..829J,2009A&A...507..105J,2010A&A...517A...4J}, whose aim is to clear up the weak lensing signal from intrinsic alignments. The idea behind this is that if we know with high accuracy the model describing the cosmology, we can then predict how, according to this model, the weak lensing spectrum must be. By subtracting the latter from the measured spectrum, the remaining part is ascribable to intrinsic ellipticities. More precisely, if the uncertainty on the covergence power spectrum is small enough to still allow for the extraction of the intrinsic ellipticity signal, then it means that we will be able, in future surveys, to notice this weak signal in the presence of the much stronger weak lensing spectrum. 

It is worth to remark, at this point, that the intrinsic alignments we consider in this work, also known as II (intrinsic-intrinsic) alignments, are not the only contaminant to the weak lensing spectrum, usually referred to as the GG (gravitational shear-gravitational shear) signal. For instance, another source of contamination are the GI (gravitational shear-intrinsic ellipticity) correlations, which might occur when the alignment produced by a dark matter halo on a closeby galaxy correlates with the shear signal that the same halo induces on a background galaxy. This effect, first suggested by \citet{2004PhRvD..70f3526H}, is difficult to remove, and is not considered in our treatment. Likewise, we do not consider complications arising from the statistical uncertainty in estimating the combined spectrum $C_\kappa(\ell)+C^\epsilon_E(\ell)$. 

The problem is therefore now to understand whether the uncertainty at which the lensing spectrum can be predicted for a given cosmology is enough to attribute a high-$\ell$ excess in the ellipticity spectrum to intrinsic alignments. In order to quantify the uncertainty on the covergence spectrum $\Delta C_{\kappa}(\ell)$, we consider a multivariate Gaussian likelihood for the cosmological parameters, draw from this distribution simultaneously a sample of five parameters, and compute for those the weak lensing spectrum. This gives us a bundle of spectra around the fiducial spectrum $C_\kappa(\ell|x_\mu^\mathrm{fid})$, and allows us to define the uncertainty as the standard deviation:
\begin{equation}
\Delta C_\kappa(\ell)^2 = 
\frac{1}{n}\sum_{i=1}^{n}\left[C_\kappa(\ell|x_\mu^{(i)}) - \bra C_\kappa(\ell|x_\mu^\mathrm{fid})\ket\right]^2,
\end{equation}
at each multipole $\ell$ where the index $i$ runs over the samples $x_\mu^{(i)}$ of parameter sets drawn from the multivariate Gaussian likelihood. In short, this sampling of a parameter set and measuring the variance of the resulting spectra is a method of propagating the statistical parameter uncertainty described by the likelihood to an error-tube around $C_\kappa(\ell)$ reflecting the prediction uncertainty in the spectrum. This allows now to compare the magnitude of the intrinsic ellipticity spectrum $C^\epsilon_E(\ell)$ to this uncertainty and to quantify the significance.

\begin{figure}
\vspace{0.5cm}
\resizebox{\hsize}{!}{\includegraphics{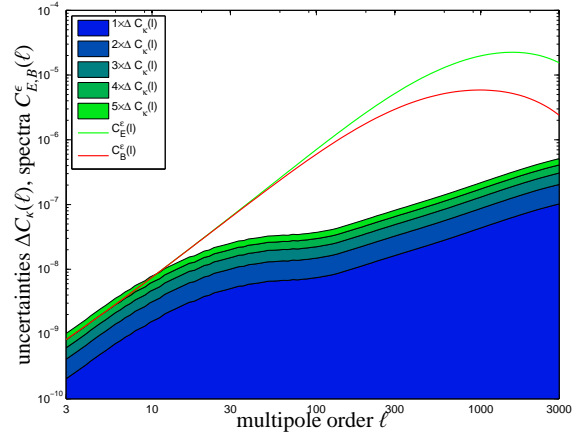}}
\vspace{0.5cm}
\caption{Uncertainties $\Delta C_\kappa(\ell)$ (shaded area) in the prediction of the nonlinear weak lensing spectra $C_\kappa(\ell)$ from drawing samples for $\Omega_m$, $\sigma_8$, $h$, $n_s$ and $w$ from a Gaussian parameter likelihood, for which we use a prior on the $w$CDM-model combining baryon acoustic oscillations, lensing and the CMB (EUCLID 10-bin BAO spectra, EUCLID 10-bin weak lensing spectra and PLANCK CMB temperature and polarisation spectra). The uncertainty is compared to the ellipticity spectra $C_E^\epsilon(\ell)$ (green line) and $C_B^\epsilon(\ell)$ (red line) determined with the CNPT-model.}
\label{fig_removal}
\end{figure}

Fig.~\ref{fig_removal} shows the uncertainties $n\times\Delta C_\kappa(\ell)$ with $n=1\dots5$ obtained by using this technique. The likelihood from which samples on the $w$CDM-parameter set including $\Omega_m$, $\sigma_8$, $h$, $n_s$ and $w$ were drawn is the one describing the knowledge on the cosmological parameters if EUCLID's measurement of baryon acoustic oscillations, EUCLID's weak lensing data (both in a 10-bin tomographic measurement) and PLANCK's CMB data were present, with a theoretical prior on spatial flatness. The PLANCK-likelihood was marginalised over the baryon density $\Omega_b$. The error tube $\Delta C_\kappa(\ell)$ can be clearly separated into two multipole ranges, the first region where the linear CDM-spectrum is dominating in the generation of the weak lensing spectrum, and the second region where the nonlinear enhancement of $P(k)$ is important and where the error tube is much wider.

With such small uncertainties and with the choice of a cosmological model with low complexity the predictive uncertainty on $C_\kappa(\ell)$ is much smaller than the amplitude of the intrinsic alignments from multipoles of $\ell\simeq30$ on, $\Delta C_\kappa(\ell)\ll C^\epsilon_E(\ell)$. We verified that the PLANCK CMB likelihood alone would not be sufficient for extracting the intrinsic ellipticity spectrum. Likewise, a more complex model with a larger number of parameters would have much larger uncertainties, $\Delta C_\kappa(\ell)\gg C^\epsilon_E(\ell)$ for most of the multipole range.

\section{Summary}\label{sect_summary}
Subject of this paper are the statistical properties of intrinsic, angular momentum induced ellipticity alignments, and their dependence on the cosmological parameter set, in comparison to ellipticity correlations induced by weak gravitational lensing. We carry out our computations with the EUCLID ellipticity data sample in mind, and use the projected EUCLID galaxy redshift distribution and shape noise  for making forecasts.

\begin{enumerate}
\item{We base our predictions for the spectra $C_E^\epsilon(\ell)$ and $C_B^\epsilon(\ell)$ describing fluctuations in the ellipticity field on physical models for angular momentum correlations in the large-scale structure \citep{2001ApJ...559..552C, 2002MNRAS.332..788M}. The two models under consideration link the angular momentum field to the tidal shear field, and model the ellipticity of a galaxy by assuming that the galactic disk is formed perpendicular to the host halo's angular momentum direction. The two models differ in describing these physical processes in configuration space versus Fourier space, and use different normalisations. For comparability, we have normalised the MWK-model such that it displays the same amplitudes as the CNPT-model on large angular scales. The CNPT-model in turn uses 3 parameters, which are the mass-scale of the galaxies, imposed by an Gaussian filter acting on the CDM-spectrum $P(k)$, a misalignment parameter $a$, which is determined to have the numerical value $a\simeq0.25$ in numerical simulations \citep{2001ApJ...555..106L} and finally the disk thickness parameter $\alpha$, which has been measured to be $\alpha\simeq0.75$ in the APM-galaxy sample \citep{2001ApJ...559..552C}.}
\item{Computing the ellipticity spectra $C_E^\epsilon(\ell)$ and $C_B^\epsilon(\ell)$ for the galaxy sample of EUCLID from both models yields spectra which are constant on large angular scales and drop off exponentially on small scales, while the $E$-mode spectrum is larger by about an order of magnitude compared to the $B$-mode spectrum on multipoles of $\ell\simeq1000$. By themselves, the spectrum $C_E^\epsilon(\ell)$ would be significantly larger than the linear weak lensing convergence spectrum $C_\kappa(\ell)$, which is comparable in amplitude to the spectrum $C_B^\epsilon(\ell)$ on these multipoles. Nonlinear structure formation, however, increases the variance of the cosmic density field strongly, such that intrinsic ellipticities contribute only $\sim20\%$ to the total variance of the ellipticity field at $\ell=1000$. Aperture weighted variances give a similar impression: The averaged shear and the aperture mass of the nonlinear weak lensing convergence dominate the variance on all relevant scales, and intrinsic alignments are smaller by at least a factor of two in this observable.}
\item{Investigating the dependence of the ellipticity spectra on cosmological parameters gives a result very different compared to other cosmological probes. Due to the dependence of the angular momentum field on the angular momentum direction and not the magnitude, $\sigma_8$ is entirely replaced by the misalignment parameter $a$. $n_s$, $\Omega_m$ and $h$ determine the CDM spectrum $P(k)$ (the latter two by fixing the shape parameter $\Gamma$) and $\Omega_m$ is of course appearing in the conversion between comoving distance and redshift at the stage of applying the Limber equation. Computing the derivatives $\partial C_E^\epsilon(\ell)/\partial x_\mu$ and $\partial C_B^\epsilon(\ell)/\partial x_\mu$ of the spectra with respect to the cosmological parameters and expressed in units of their covariance suggests that the parameters $a$ and $\Omega_m$ are the ones most important for intrinsic alignments, with only minor dependences on the dark energy equation of state $w$ and the Hubble-parameter $h$. At the same time $\Omega_m$ and $\sigma_8$ are the ones best constrained by lensing, so that it is suggestive to expect the largest estimation biases in those two parameters, if the intrinsic alignments are not properly removed or modelled.}
\item{In the next step we quantified the likelihood $\mathcal{L}(\Omega_m,a,w)$ of the intrinsic alignment spectrum $C_E^\epsilon(\ell)$ if lensing was not present. One could expect this likelihood to have non-Gaussian contributions because of the nonlinearities present in an angular momentum-based alignment model: firstly, the angular momentum depends on the quadratic tidal shear and the ellipticity depends on the squared angular momentum direction. For a resticted multipole $\ell_\mathrm{max}\lsim100$ range one can see clear deviations from Gaussianity, which quickly vanish if the multipole range is extended. The misalignment parameter $a$, which describes the normalisation of the spectra, is described by a Gaussian likelihood and enters as a prefactor.}
\item{We compute estimation biases on the $w$CDM parameter set if intrinsic alignments are not removed from the ellipticity spectrum, i.e. if the data is in reality described by $C_\kappa(\ell)+C_E^\epsilon(\ell)$ and wrongly fitted by $C_\kappa(\ell)$ only. The strongest biases are present in $\Omega_m$, which is measured to low and $\sigma_8$, which is estimated too high, both at the level of $\sim2\sigma$, making the estimation biases significant. Interestingly, the dark energy equation of state is almost unbiased, indicating that dark energy investigations are not directly affected by intrinsic alignments. Changing the magnitude of the intrinsic alignment contamination by increasing the misalignment parameter $a$ shows an monotonic increase of the estimation biases for all parameters except $n_s$ where the estimation bias saturates at $a\simeq0.5$ and drops for higher amplitudes. Clearly, these results demand a good external prior on $a$, either from independent measurements or from numerical simulations.}
\item{Finally we investigate if the weak lensing convergence spectrum $C_\kappa(\ell)$ can be predicted precisely enough such that a deviation can be attributed to a contribution $C_E^\epsilon(\ell)$. For this purpose, we develop a technique for propagating the uncertainty in the set of cosmological parameters to the variance $\Delta C_\kappa(\ell)^2$ around $C_\kappa(\ell)$ for the fiducial cosmology. Comparing this uncertainty with the amplitudes $C_E^\epsilon(\ell)$ suggests that it should be measurable at high multipoles. In this process we used a Gaussian likelihood $\mathcal{L}$ on a standard $w$CDM-cosmology reflecting the knowledge on the cosmological parameters from EUCLID's BAO- and weak lensing spectra and from the temperature and polarisation spectra of the cosmic microwave background measured by PLANCK.}
\end{enumerate}

We plan to extend our research to the intrinsic alignment contamination of tomographic weak lensing data, and to include cross-correlations between the weak lensing shear and the intrinsic ellipticity field, the so-called GI-alignments, which we aim to derive from angular momentum-based alignment models and which enter the ellipticity spectra at higher order. These GI-alignments are challenging to describe as they introduce ellipticity correlations across tomography bins. Additionally, we aim to include a linear alignment model for elliptical galaxies and to work with a proper morphological mix of ellipticities, working towards a more complete physical description of alignments in tomographic weak lensing data.

\section*{Acknowledgements}
Fisher-matrices for the description of parameter constraints forecast for EUCLID's tomographic weak lensing survey and EUCLID's BAO measurements are taken from the iCosmo resource, and the corresponding Fisher-matrix for PLANCK's CMB-spectrum has been kindly provided by Maik Weber and Youness Ayaita. We would like to thank Matthias Bartelmann, Angelos F. Kalovidouris and Vanessa B{\"o}hm for their support and their suggestions, and are grateful for feedback from Ana{\"i}s Rassat and Lukas Hollenstein. FC receives funding through DFG's Schwerpunktprogramm SPP1177, PhMM is supported by the Graduate Academy, Heidelberg and VMB is funded by the German Academic Exchange Service. Both FC and PhMM acknowledge support from the International Max Planck Research School for Astronomy and Cosmic Physics in Heidelberg and from the Graduate School of Fundamental Physics. BMS's work was supported by the German Research Foundation (DFG) within the framework of the excellence initiative through the Heidelberg Graduate School of Fundamental Physics.

\bibliography{bibtex/aamnem,bibtex/references}
\bibliographystyle{mn2e}

\appendix

\bsp

\label{lastpage}

\end{document}